\documentclass[journal]{IEEEtran}
\usepackage{graphicx}
\usepackage{url}%

\usepackage{cite}
\ifCLASSINFOpdf
\else
\fi
\usepackage{subfigure}

\def\Fig#1{Fig.~\ref{fig:#1}}

\def\Tab#1{Table~\ref{tab:#1}}

\newcommand*{\rom}[1]{\expandafter\@slowromancap\romannumeral #1@}
\makeatother
\begin{document}
\title{Electroencephalography based Classification of Long-term Stress using Psychological Labeling}

\author{Sanay~Muhammad~Umar~Saeed,~Syed~Muhammad~Anwar,~Humaira~Khalid,~Muhammad~Majid,~Ulas~Bagci% <-this % stops a space
%\thanks{A. Mehreen, M. Haseeb and M. Obaid are with the Department of Electrical Engineering, University of Engineering and Technology, Taxila}% <-this % stops a space
\thanks{S.M.U.Saeed and M.Majid are with the Department of Computer Engineering, University of Engineering and Technology, Taxila, Pakistan}% <-this % stops a space
\thanks{S.M.Anwar is with the Department of Software Engineering, University of Engineering and Technology, Taxila, Pakistan}% <-this % stops a space
\thanks{H. Khalid is with the Department of Psychology, Benazir Bhutto International Hospital, Rawalpindi, Pakistan} % <-this % stops a space
\thanks{U. Bagci and S.M.Anwar are with Center for Research in Computer Vision (CRCV), University of Central Florida, Florida, USA (s.anwar@knights.ucf.edu)}}% <-this % stops a space


\markboth{IEEE Journal of Biomedical and Health Informatics}%
{Mehreen \MakeLowercase{\textit{et al.}}: Bare Demo of IEEEtran.cls for IEEE Journals}



\maketitle

\begin{abstract} Stress research is a rapidly emerging area in the field of electroencephalography (EEG) based signal processing. %There are methods reported in literature to quantify human acute stress in response to induced stressors using EEG signal recordings. In comparison, the classification of long-term or chronic stress using EEG has not been widely assessed. 
The use of EEG as an objective measure for cost effective and personalized stress management becomes important in particular situations such as the non-availability of mental health facilities. In this study, long-term stress is classified using baseline EEG signal recordings. The labelling for the stress and control groups is performed using two methods (i) the perceived stress scale score and (ii) expert evaluation. The frequency domain features are extracted from five-channel EEG recordings in addition to the frontal and temporal alpha and beta asymmetries. The alpha asymmetry is computed from four channels and used as a feature. Feature selection is also performed using a t-test to identify statistically significant features for both stress and control groups. %Statistically differentiating features are not found in stress and control group, when labels are assigned by the PSS method, whereas beta and gamma waves from $AF3$ channel are found to be statistically different in stress and control group along with overall alpha ratio, when subjects are labelled using expert evaluation.% 
We found that support vector machine is best suited to classify long-term human stress when used with alpha asymmetry as a feature. It is observed that expert evaluation based labelling method has improved the classification accuracy up to 85.20\%. Based on these results, it is concluded that alpha asymmetry may be used as a potential bio-marker for stress classification, when labels are assigned using expert evaluation. 
\end{abstract}
\begin{IEEEkeywords}
Long-term stress, electroencephalography (EEG), machine learning, perceived stress scale, expert evaluation.  
\end{IEEEkeywords}

% \fi
\IEEEpeerreviewmaketitle

\section{Introduction}%[Paragraph related to the journal]

\IEEEPARstart{T}{he} response of human body to a demand for change is considered as stress\cite{selye1965stress}. A balance exists between the sympathetic and parasympathetic arms of the autonomic nervous system in healthy people. A fight-or-flight response is invoked when there is an exposure to a threatening situation. Daily routine stress does not pose any danger to life but, the fight-or-flight response may still be invoked. A persistence of this short-term stress for a longer duration can cause long lasting effects on the neurology of an individual and may give rise to depression \cite{heim2002neurobiology}. Long-term stress is a better predictor of depressive symptoms as compared to short-term stress \cite{mcgonagle1990chronic}. Long term stress is considered as a risk factor for many health conditions such as cardiovascular diseases \cite{cohen2007psychological,steptoe2012stress}. 

The prevention of the onset of depression requires a timely detection of long-term stress symptoms. Conventional psychological methods and analysis of hormones such as cortisol and alpha-amylase are widely used in long-term stress studies \cite{van2004can}. These methods are practical but they are affected by various factors such as culture, language, and objectivity. For instance, Perceived Stress Scale (PSS) is a widely used questionnaire to measure the level of chronic stress, validated extensively across diverse samples \cite{hammen2014measurement}. In general, a self-administered checklist cannot equal the precision of an interviewer trained to elicit aspects of events critical to examine stress. Such interviews have shown to provide substantially better information in comparison to relatively unassisted self-reporting mechanisms \cite{sobell1990procedure}. The respondents have been found to report minor or positive events in response to questions designed to elicit negative and undesirable events \cite{mcquaid1992toward}. 

Psychological methods alone are not enough to assess stress related conditions \cite{peng2013method}. Stress can be quantified objectively from bio-markers like electroencephalography (EEG), galvanic skin response, and electrocardiography \cite{zheng2015biosignal}. EEG is one of the most common source of information for studying brain function \cite{8664170} \cite{asif2019human} \cite{saeed2018selection} \cite{raheel2018emotion} \cite{anwar2018game}. The oscillations generated by the variation of electric potential in the brain are recorded using low resistance electrodes placed on the human scalp \cite{sanei2007eeg}. It is a widely used non-invasive method due to its excellent temporal resolution, ease of use, and low cost. EEG signals are categorized by their frequency bands including delta, theta, alpha, beta and gamma. Each frequency band can be used as a discriminating feature for different brain states \cite{al2018towards}. There are methods reported in literature to quantify human acute stress in response to induced stressors using EEG signal recordings. In comparison, the classification of long-term or chronic stress using EEG has not been widely assessed.

%Hemispheric specialization is a major concern in neuro-physiological research. Generally, a healthy brain at rest has a fairly balanced level of activity in each hemisphere \cite{fisch1999fisch}. Left hemisphere is associated with the processing of positive emotions, while right hemisphere is associated with the processing of negative emotions \cite{davidson2004does}. Moreover, the extent of this asymmetry has been suggested to vary under conditions of chronic stress \cite{papousek2002covariations}. Frontal asymmetry is highly related to post-traumatic stress disorder \cite{lobo2015eeg}. The results in \cite{goncharova1990changes}, have shown that the major depression disorder (MDD) group is significantly right lateralized relative to controls, and both MDD and post-traumatic stress disorder (PTSD) displayed more left- than right-frontal activity.%

In this study, the problem of long-term human stress recognition is addressed by using PSS labels and expert evaluation. Two groups of participants were considered including stressed and control group. A total of forty five different features were extracted from EEG signals in frequency domain to classify these two groups. Discriminating features were selected using a statistical significance test. Five different classifiers including support vector machine (SVM), Naive Bayes (NB), K-nearest neighbour (KNN), logistic regression (LR) and multi-layer perceptron (MLP) were used to classify human stress using the selected features. The summary of our contributions in this study is as follows,

\begin{enumerate}
\item EEG signals were acquired from 33 participants in closed eye condition using a five-channel EEG headset and three frequency domain features were found statistically significant in stress and control groups. 
\item To the best of our knowledge, it is for the first time that the stress level of participants was labelled by a psychology expert in an EEG based study.
\item The machine learning based method suites well to long-term human stress classification and gives better performance using psychological expert labeling.
\end{enumerate}

The rest of the paper is organized as follows: Section \ref{RW}, describes the related work, Section \ref{Meth} presents the proposed methodology including data collection, feature extraction, and classification algorithms. Section \ref{ReD} presents the results and a comparison with previously reported studies. Finally, the conclusion of the study is given in section \ref{sec:conc}.

\section{Related Work} \label{RW}
Hemispheric specialization is a major concern in neuro-physiological research. Generally, a healthy brain at rest has a fairly balanced level of activity in both hemispheres of brain \cite{fisch1999fisch}. The left hemisphere is associated with the processing of positive emotions, while the right hemisphere is associated with the processing of negative emotions \cite{davidson2004does}. The extent of asymmetry has been suggested to vary under conditions of chronic stress \cite{papousek2002covariations}. Frontal asymmetry is highly related to post-traumatic stress disorder (PTSD) \cite{lobo2015eeg}. The results in \cite{goncharova1990changes}, have shown that major depression disorder (MDD) group is significantly right lateralized relative to controls, and both MDD and PTSD displayed more left- than right-frontal activity.

\begin{figure*}[!ht]
\begin{center}
\includegraphics[width = 140mm]{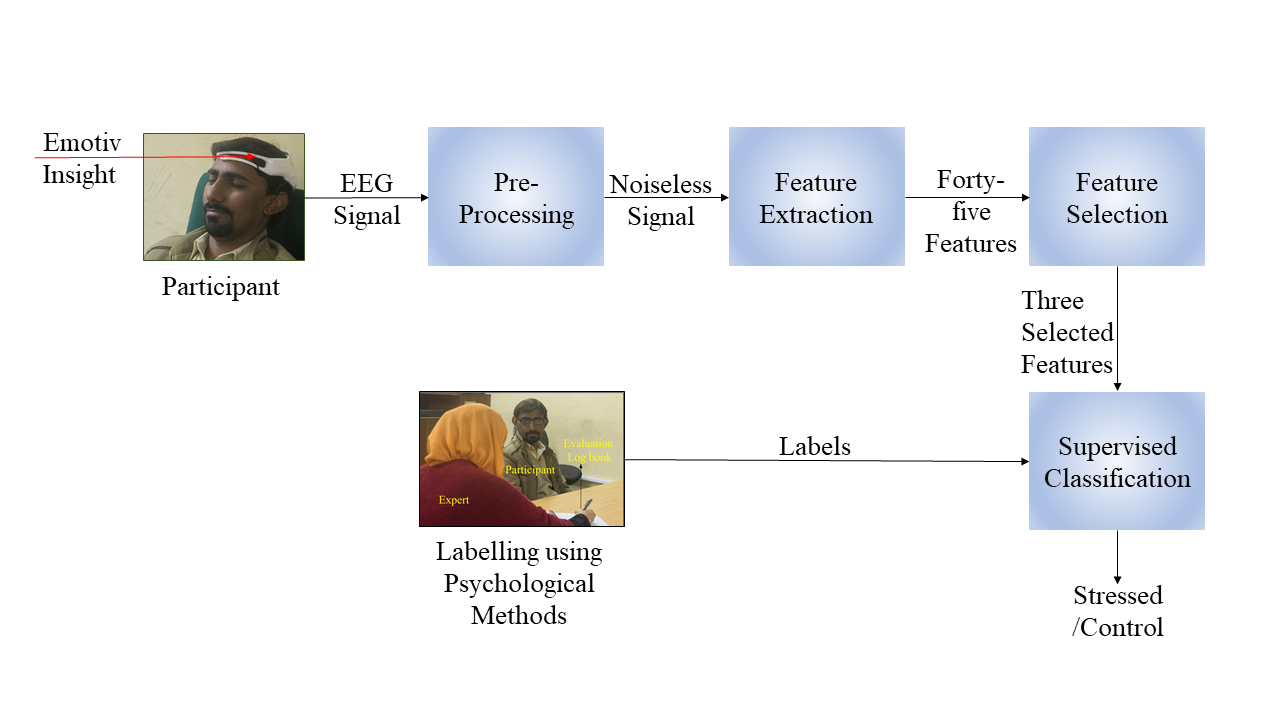}
\caption{The proposed methodology for long-term human stress classification.}
\label{fig:fig1}
\end{center}
\end{figure*}

Recently, the feasibility of using EEG in classifying multilevel mental stress has been demonstrated \cite{al2018towards}, where alpha rhythm at the right pre-frontal cortex was suggested as a suitable bio-marker. A machine learning framework using EEG signals was proposed in \cite{subhani2017machine}, where stress was induced by using the Montreal imaging stress task (MIST), and SVM, NB, and LR classifiers were used to classify the stress level of participants. The EEG of participants in resting-state was recorded under negative, positive and neutral stimulus using soundtracks from the international affective digitized sounds (IADS-2) dataset  \cite{cai2018pervasive}. Stress detection based on frontal alpha asymmetry was performed using the DEAP dataset, and classification was performed using SVM, KNN and fuzzy KNN \cite{baghdadi2017efficient}. In \cite{aspinall2015urban}, mobile EEG was used to assess stress in humans using EMOTIV EPOC headset in an out-of-lab environment. In an EEG based study, $11$ participants were analyzed for the identification of long-term stress \cite{peng2013method}, including 7 mothers of children with mental disability (stress group) and 4 mothers of healthy children (control group). %\textbf{Similarly, baseline EEG was used to test the relation of theta and beta ratio to self-reported trait attention based control \cite{putman2014eeg}.}[meaning not clear] 

A variant of trier social stress task (TSST) was used to assess stress involving forty nine participants \cite{dusing2016relative}. Samples of the salivary cortisol and baseline EEG based alpha asymmetry were assessed before and after performing TSST. The frontal and parietal alpha asymmetry was used to classify depression in elderly people \cite{kaiser2018electroencephalogram}. The correlation between frontal and parietal alpha asymmetry, the geriatric depression scale, and the mini mental state examination were analyzed. A high beta activity at the frontal and occipital lobes was observed on the visual input of negative images \cite{seo2010stress}. The frontal theta activity was shown to decrease due to a stressful mental arithmetic task \cite{gartner2015frontal}. %On the other hand, high delta waves were observed, while solving a difficult mental arithmetic task \cite{harmony1996eeg}. 
In \cite{saeed2017quantification}, low beta waves in closed eye condition were found to be a strong predictor of perceived stress, where PSS score was predicted by using multiple linear regression. The pre-frontal relative gamma power i.e., the ratio of gamma band and slow brain rhythms, was proposed as a bio marker for identification of stress \cite{minguillon2016stress}. 

The related studies presented here can be grouped as either short-term or long-term stress assessment. Short-term stress is measured using a stress eliciting task, while long-term stress is measured without performing any additional mental task. Different techniques have been adopted to measure stress, but most of these techniques require human intervention. Among different physiological measures, EEG has the potential to be used as a measure of stress in daily life. This is due to the fact that EEG headsets are becoming commercially available for observing brain activity in an easy to wear and cost effective manner. The proposed study uses EEG signals acquired with a commercially available EEG headset to identify baseline or long-term stress without relying on stress inducing tasks. 

\section{Methodology}\label{Meth}
 %In data acquisition step, brain activity is recorded from human scalp using EEG headset in a predefined condition. Recorded EEG signals are made noise free in the pre-processing stage, after which the key features are extracted from EEG signals. Features that are more representative of the underlying data are selected by using feature selection methods. Finally, extracted features are used by classification algorithms to classify the stress level of a human.
\begin{figure*}[!t]
\begin{center}
\includegraphics[width = 170mm]{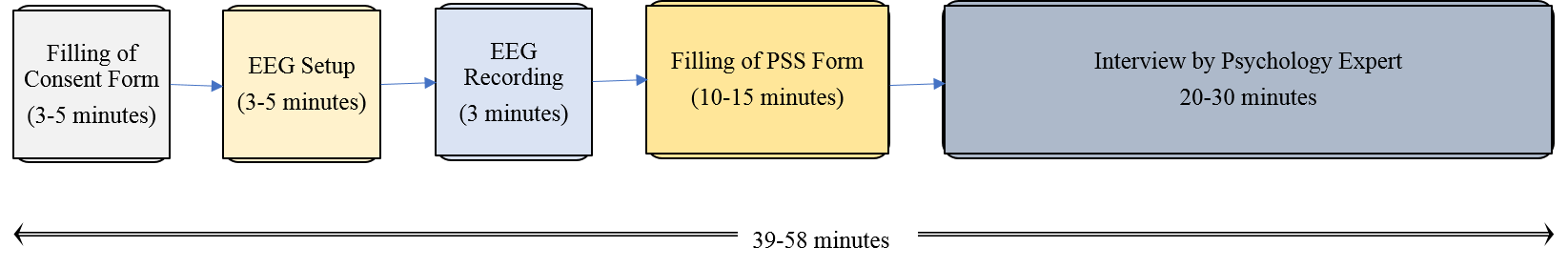}
\caption{The sequence of events during the data acquisition process.}
\label{fig:figTimeFlow}
\end{center}
\end{figure*}
%The methodology followed in the proposed study is shown in \Fig{fig1}, where 
We devised a supervised model for the classification of human stress (\Fig{fig1}). A total of thirty three volunteers participated in this study. The baseline EEG data for each participant were acquired using an EMOITV Insight headset (\url{https://www.emotiv.com/insight/}) in a closed eye condition for three minutes. After EEG signal recording, participants were asked to fill in the PSS-10 questionnaire followed by an interview with the psychology expert. The average time for the interview was $25$ minutes. Based on the PSS scores and interview, the psychology expert grouped each participant in either stress or control group. 

The recorded EEG signals were made noise free in the pre-processing stage. Neuro-physiological features including alpha ($\alpha$), low beta ($\beta_{l}$), beta ($\beta$), gamma ($\gamma$), delta ($\delta$), theta ($\theta$), and relative gamma (RG) power were extracted from the signals at each electrode. Frontal and temporal alpha and beta asymmetries, and alpha asymmetry was calculated from these features. Five supervised machine learning algorithms (SVM, NB, KNN, LR and MLP) were used to classify human stress. Two different labeling methods were used, including the perceived stress scale and expert evaluation, where the PSS and interview scores were simultaneously used. A detailed description of these methods is presented in the following subsections. 

\subsection{Data Acquisition} All EEG recordings are performed in a noise free lab using EMOTIV Insight headset, which records brainwaves and provides advanced electronics that are optimized to produce clean and robust signals. Its data transmission rate is $128$ samples per second, which provides the ability to perform an in-depth analysis on the brain activity. It has a minimum voltage resolution of $0.51$ volts least significant bit with $5$ EEG sensors at $AF3, AF4, T7, T8, Pz$ locations and $2$ reference electrodes. The participants are asked to close their eyes for a duration of three minutes and are instructed to keep their head still to reduce movement artifacts. Closed-eye condition is used, since correlates of long-term stress have been found in this condition in previous studies \cite{saeed2017quantification,peng2013method}. Another advantage of using the closed eye condition is the minimization of eye blink artifact. EEG signal acquisition is performed using EMOTIV Xavier TestBench v.3.1.21. EEG signals are recorded from the scalp of participants, while they are seated in a comfortable chair. The flow of events during the data acquisition process is shown in \Fig{figTimeFlow}  

\subsection{Pre-processing}
The EEG signals recorded from the scalp contain noise due to external interference. Before feature extraction, noise is removed from the signals for better classification results. The continuous offset is removed from EEG signals by computing an average of each channel over the baseline and by subtracting it from the channel. This removes the baseline value of each sensor i.e., the continuous offset that is added permanently on top of the EEG recordings. For reducing muscular artifacts, participants are instructed to minimize their head movements during the EEG acquisition. In closed eye condition, blink artifacts are also found to be minimal. EMOTIV Insight has a frequency response of $1-43$ Hz, which makes the signal noise free from AC line interference at $50$ Hz.

\subsection{Feature Extraction and Selection}
Neural oscillatory features are widely used in literature for EEG based classification systems. EEG signals are decomposed into different frequency bands. The Welch method is used to extract power spectral densities with a window length of $128$ samples with $50$ percent overlap. For feature extraction, power spectral densities of different neural oscillations namely, delta ($1-3$ Hz), theta ($4-7$ Hz), alpha ($8-12$ Hz), beta ($13-30$ Hz), gamma ($25-43$ Hz), slow ($4-13$ Hz), and low beta ($13-17$ Hz) are computed from each channel. Relative gamma waves are computed by taking the ratio of slow and gamma waves. Eight features from each of the five channels adds up to forty features. Moreover, five alpha and beta asymmetries are calculated, giving a total of forty-five neural oscillatory features. The alpha asymmetries are calculated using the following equations,

\begin{equation}\label{eq:eq1}
\alpha_f=\frac{\alpha_{AF4}-\alpha_{AF3}}{\alpha_{AF3}+\alpha_{AF4}},
\end{equation}
\begin{equation}\label{eq:eq2}
\alpha_t=\frac{\alpha_{T8}-\alpha_{T7}}{\alpha_{T8}+\alpha_{T7}},
\end{equation}
\begin{equation}\label{eq:eq5}
\alpha_a=\alpha_f+\alpha_t,
\end{equation}
where $\alpha_f, \alpha_t$ and $\alpha_a$ represents the frontal alpha, temporal alpha, and alpha asymmetry respectively and $\alpha_{channel}$ represents the alpha power spectral density of the frontal and temporal EEG channels. Similarly, the frontal and temporal beta asymmetries are calculated using, 
\begin{equation}\label{eq:eq3}
\beta_f=\frac{\beta_{AF4}-\beta_{AF3}}{\beta_{AF3}+\beta_{AF4}},
\end{equation}
\begin{equation}\label{eq:eq4}
\beta_t=\frac{\beta_{T8}-\beta_{T7}}{\beta_{T8}+\beta_{T7}},
\end{equation}
where $\beta_f$, and $\beta_t$ represents the frontal and temporal beta asymmetries and $\beta_{channel}$ represents the beta power spectral densities for the frontal and temporal EEG channels. Features are selected using a t-test for determining the statistical significance of features in stress and control group. A lower p-value returned by the t-test shows that the feature is significantly discriminating in stress and control group.

%\begin{figure}[!t]
%\begin{center}
%\includegraphics[width = 80mm]{PSSlabel.eps}
%\caption{The steps used in the PSS based labeling method for stress and control group.}
%\label{fig:fig3}
%\end{center}
%\end{figure}

%\begin{figure}[!t]
%\begin{center}
%\includegraphics[width = 80mm]{expert.eps}
%\caption{The steps used in the expert evaluation based labeling method for stress and control group.}
%\label{fig:fig4}
%\end{center}
%\end{figure}

%\begin{figure}[!t]
%\begin{center}
%\includegraphics[width = 60mm]{log.png}
%\caption{A participant while being interviewed by the psychology %expert (The picture is used with an informed consent of the participant).}
%\label{fig:fig2}
%\end{center}
%\end{figure}

\subsection{Subject labelling}
The proposed method uses two types of labelling for supervised classification. PSS-10 is used for the questionnaire based labelling method to subjectively evaluate the stress of participants. %A block diagram of labelling method using PSS scores is shown in \Fig{fig3}. 
This questionnaire consists of ten questions. Each question asks the subject about the frequency of stressful events that have occurred during a period covering the last thirty days. The response for each question is on a scale of $0$ to $4$, where $0$ represents that the event never occurred and $4$ represents a frequent occurrence. The total PSS-10 score for each participant has a range between $0$ and $40$. The participants are divided in two groups i.e., the control and stress group, using the PSS score. A threshold is selected for this purpose, which is given by the following equation, 

\begin{equation}\label{eq:eq8}
T_p=\mu\pm\frac{\sigma}{2},
\end{equation}
where, $T_p$ is threshold of PSS score, $\mu$ is the mean and $\sigma$ is standard deviation of the PSS scores. 

%\subsection{Expert evaluation based labeling}
The psychologist assigned labels for the stress and control groups after an expert evaluation based on the interview and PSS scores. During the interview the expert investigated about the physical, emotional, behavioral and cognitive symptoms of stress. Physical symptoms included aches or pain, diarrhea or constipation, nausea, dizziness, chest pain and rapid heart rate. Emotional symptoms of stress included depression, anxiety, moodiness, irritability, overwhelming feel and loneliness.  Behavioral and cognitive symptoms included memory problems, inability to concentrate, poor judgment, negativity, racing thoughts and constant worrying. The interviews are conducted by the psychologist, affiliated with a public sector hospital. %A participant can be seen in \Fig{fig2} during the interview process, where the expert is making a log of symptoms reported by the participants for stress assessment.

\begin{figure*}[!ht]
\begin{center}
\includegraphics[width = 120mm]{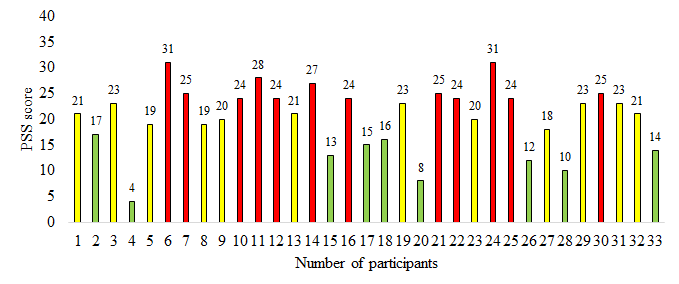}
\caption{A histogram of PSS scores for participants showing labels assigned using the PSS based labelling method (green: control group, red: stress group, yellow: neutral).}
\label{fig:fig5}
\end{center}
\end{figure*}

\begin{table*}[!t]
\centering
\caption{Results for t-test on various neural oscillations including PSS and expert based labelling methods.} 
\label{PSSt}
\scalebox{1}{
\begin{tabular}{c |c |c |c |c |c| c| c| c| c}
\hline\hline
\textbf{Labelling} & & \multicolumn{8}{|c}{\textbf{Neural Oscillations}}   \\  
\textbf{Method} & \textbf{Channel} & \textbf{delta} ($\delta$) & \textbf{theta} ($\theta$) & \textbf{slow}  & \textbf{alpha} ($\alpha$) & $\beta_{l}$ & \textbf{beta} ($\beta$)  & \textbf{gamma} ($\gamma$) & \textbf{RG} \\
%Method & & & & & & & & & \\
\hline
& AF3       & 0.12 & 0.09 & 0.13 & 0.28 & 0.30 & 0.21 & 0.32 & 0.53 \\
& T7       & 0.89 & 0.81 & 0.61 & 0.21 & 0.58 & 0.85 & 0.52 & 0.36 \\
\textbf{PSS} & Pz       & 0.15 & 0.16 & 0.16 & 0.19 & 0.29 & 0.46 & 0.64 & 0.30 \\
& T8       & 0.89 & 0.97 & 0.95 & 0.87 & 0.49 & 0.97 & 0.90 & 0.26 \\
& AF4       & 0.14 & 0.12 & 0.13 & 0.22 & 0.20 & 0.15 & 0.23 & 0.79\\
\hline
& AF3      & 0.65 & 0.50 & 0.51 & 0.08 & 0.95 & \textbf{0.04}   & \textbf{0.03}    & 0.23 \\
& T7       & 0.92 & 0.60 & 0.51 & 0.15 & 0.99 & 0.42   & 0.54 & 0.99 \\
\textbf{Expert} & Pz       & 0.91 & 0.89 & 0.90 & 0.90 & 0.93 & 0.69   & 0.34 & 0.40 \\
& T8       & 0.54 & 0.51 & 0.55 & 0.48 & 0.85 & 0.96   & 0.85 & 0.56 \\
& AF4      & 0.11 & 0.12 & 0.12 & 0.35 & 0.25 & 0.21   & 0.28    & 0.61\\ \\
\hline
\hline
\end{tabular}}
\end{table*}

\subsection{Stress Classification}
In this study, five different types of classifiers are used for classification, which are described in the following subsections.
\subsubsection{Support Vector Machine}
A support vector machine uses the statistical learning theory, which is based on the principle of structural risk minimization. An SVM selects a hyper-plane, which separates the feature space in to control and stress group according to the labels provided. The SVM is a highly efficient classifier and is used widely for stress classification in EEG based studies \cite{subhani2017machine,al2018towards}. The use of SVM reduces the risk of data over-fitting and provides good generalization performance. 

\subsubsection{The Naive Bayes}
Naive Bayes is a probabilistic classifier based on Bayes theorem. It uses maximum posterior hypothesis of statistics and works well for high dimensional input data. It is a nonlinear classifier and gives good results in real world problems. In addition, the Naive Bayes classifier requires a small amount of training data to approximate the statistical parameters \cite{kotsiantis2007supervised}.

\subsubsection{K-nearest Neighbors}
KNN is an instance based learning classifier, where training instances are stored in their original form. A distance function is used to determine the  member of the training set, which is nearest to a test example and used to predict the class. The distance function is easily determined if the attributes are numeric. Most instance-based classifiers use Euclidean distance for distance calculation. The distance between an instance with attribute values $a_1,a_2, . . . , a_n$
(where n is the number of attributes) and $b_1, b_2,...,b_n$ is defined as, 

\begin{equation}\label{eq:eq2}
D_g=\sqrt{(a_k-b_k)^2}.
\end{equation}

\subsubsection{Logistic Regression}
The logistic regression algorithm guards against over-fitting by penalizing large coefficients. The output is set to one for training instances belonging to the class and zero otherwise. Logistic regression builds a linear model based on a transformed target variable, where a transformation function converts a non-linear function to a linear function.

\subsubsection{Multi-layer Perceptron}
In a multi-layer perceptron structure, transfer functions are used for mapping inputs to the output. These functions include sigmoid function, rectified linear unit and hyperbolic tangent. The classifier uses back-propagation to classify instances. Multi-layer perceptrons are trained by minimizing the squared error of the network output, essentially treating it as an estimate of the class probability, which is given by the following equation,
\begin{equation}\label{eq:eq0}
E=\frac{1}{2}((y-f(x)^2)),
\end{equation}
where, $f(x)$ is the network prediction obtained from the output unit and $y$ is the instance class label.

\section{Results and Discussion} \label{ReD}

\subsection{Dataset}
A total of thirty three participants related to education field volunteered for this study. The participants reported no history of brain injury and they were not using any medication, which could have affected their brain activity at the time of experiment. Among these $33$ healthy participants, $20$ are male and $13$ are females ($60.6\%$ male and $39.4\%$ female). The participant's ages ranged from $18$ to $40$ years ($\mu=23.85$, $SD=5.48$). In line with the Helsinki Declaration \cite{general2014world} and the departmental ethics guidelines, all participants of the study were briefed about the research goals. In addition, a signed informed consent was obtained from each participant. 

\subsection{Performance Parameters}
For performance evaluation, the parameters used in this study include average accuracy rate, Kappa statistic, F-measure, mean absolute error (MAE), and root mean absolute error (RMAE). Accuracy is the ratio of truly classified instances over total number of instances in the recorded data. F-measure is calculated by considering the precision and recall values. The Kappa  statistic values ranges between $0$ and $1$, where $0$ represents chance level classification and $1$ means perfect classification. A value less than zero shows that the classification is worse than chance level. A 10-fold cross validation technique was used in this study, where the training data was randomly divided into ten equal parts and the process is repeated 10 times. During the process, every instance was used for testing at a time and the remaining instances were used for training of the classifier.

\subsection{Stress and Control Group}

The histogram of PSS scores acquired from the PSS questionnaire is shown in \Fig{fig5}. The green and red bars represent the PSS scores of participants belonging to the control and stress groups respectively. The yellow bars indicate the PSS scores of participants not considered in either stress or control group. Overall, for the PSS scores we have $(\mu, \sigma) = (20.4 \pm 6.14)$. %is $20.4$, and $\sigma$ is $6.14$. 
A participant with a PSS score below $17.33$ is considered to be in control group, whereas a participant with a PSS score higher than $23.47$ is categorized in the stress group. These values are calculated using the threshold criteria defined in Eq. 6. Hence $11$ participants are labelled in stress group (red bars) and $9$ in control group (green bars).  

%\subsection{Expert's evaluation based labeling}
In expert evaluation, the psychology expert considered both the PSS score and the symptoms obtained from the interview method. The expert interviewed each participant for an average duration of $25$ minutes. Out of the $33$ participants, $10$ are assigned to the stress group, while $10$ are assigned to the control group. %The details about each participant regarding gender, PSS score, the label assigned by using PSS score and the label assigned by expert is given in \Tab{participants}.

%\begin{table}[!t]
%    %\caption{Global caption}
%%\begin{minipage}{.3\linewidth}
%   %\tbl
%      \caption{Demographics of participants and their labels, A=member of stress group, B=member of control group, X=not considered for stressed and control group}
%      \label{tab:participants}
%      \centering
%      \scalebox{0.7}{
%              \begin{tabular}{c c c c c c c}    % centered columns (4 columns)
%\hline\hline %inserts double horizontal lines
%Participant Number & Gender & Age(years) & PSS label  &  Expert label&\\
%                        \hline    % inserts single horizontal line
%1  &  M & 28 & X & X\\ % inserting body of the table
%2  &  M & 29 & X & X\\
%3  &  M & 21 & X & X\\
%4  &  M & 32 & A & A\\
%5  &  F & 19 & X & A\\
%6  &  F & 18 & B & B\\
%7  &  M & 24 & B & X\\
%8  &  M & 33 & A & A\\
%9  &  M & 21 & X & B\\
%10 &  M & 22 & B & X\\
%11 &  F & 20 & B & B\\
%12 &  M & 19 & B & B\\
%13 &  M & 24 & X & A\\
%14 &  F & 20 & B & B\\
%15 &  M & 23 & A & X\\
%16 &  M & 21 & X & X\\
%17 &  F & 19 & A & A\\
%18 &  M & 25 & A & A\\
%19 &  F & 21 & X & B\\
%20&   M & 34 & A & A\\
%21&   M & 33 & B & X\\
%22&   F & 21 & B & B\\
%23&   M & 31 & X & B\\
%24&   F & 24 & B & B\\
%25&   F & 20 & X & B\\
%26&   M & 19 & A & A\\
%27&   M & 21 & X & A\\
%28&   M & 21 & A & X\\
%29&   F & 21 & X & X\\
%30&   F & 23 & B & X\\
%31&   M & 20 & X & X\\
%32&   M & 40 & X & X\\
%33&   F & 20 & A & A\\
%\hline          %inserts single line
%\end{tabular}}
%\end{minipage}%
%\end{table}

\subsection{Feature Selection using t-test}

%\begin{table}[!ht]
%\centering
%\caption{Lables by the expert}
%\label{expertt}
%\scalebox{0.7}{
%\begin{tabular}{ccccccccc}
%\hline\hline
%Channel & delta & theta & slow  & alpha & low   & beta    & gamma    & RG    \\
%\hline
%1       & 0.651 & 0.502 & 0.512 & 0.076 & 0.954 & 0.038   & 0.034    & 0.231 \\
%2       & 0.917 & 0.604 & 0.513 & 0.146 & 0.986 & 0.424   & 0.541    & 0.992 \\
%3       & 0.908 & 0.894 & 0.900 & 0.903 & 0.926 & 0.689   & 0.338    & 0.399 \\
%4       & 0.545 & 0.508 & 0.554 & 0.478 & 0.846 & 0.962   & 0.846    & 0.562 \\
%5       & 0.112 & 0.122 & 0.117 & 0.347 & 0.250 & 0.207   & 0.276    & 0.607\\
%\hline
%\end{tabular}}
%\end{table}

\begin{table}[!t]
\centering
\caption{Asymmetries in PSS and expert evaluation.}
\label{tab:ratiot}
\begin{tabular}{c|c|c|c|c|c}
\hline\hline

Labelling   & $\alpha_{t}$ & $\alpha_{f}$  & $\beta_{t}$  & $\beta_{f}$ & $\alpha_a$ \\
Method & & & & & \\       
       \hline
PSS    & 0.23 & 0.39 & 0.91 & 0.45 & 0.11 \\

Expert & 0.21 & 0.07 & 0.49 & 0.73 & \textbf{0.0005}  \\
\hline 
\end{tabular}
\end{table}

\begin{figure*}[!t]
\begin{center}
\includegraphics[width = 160mm]{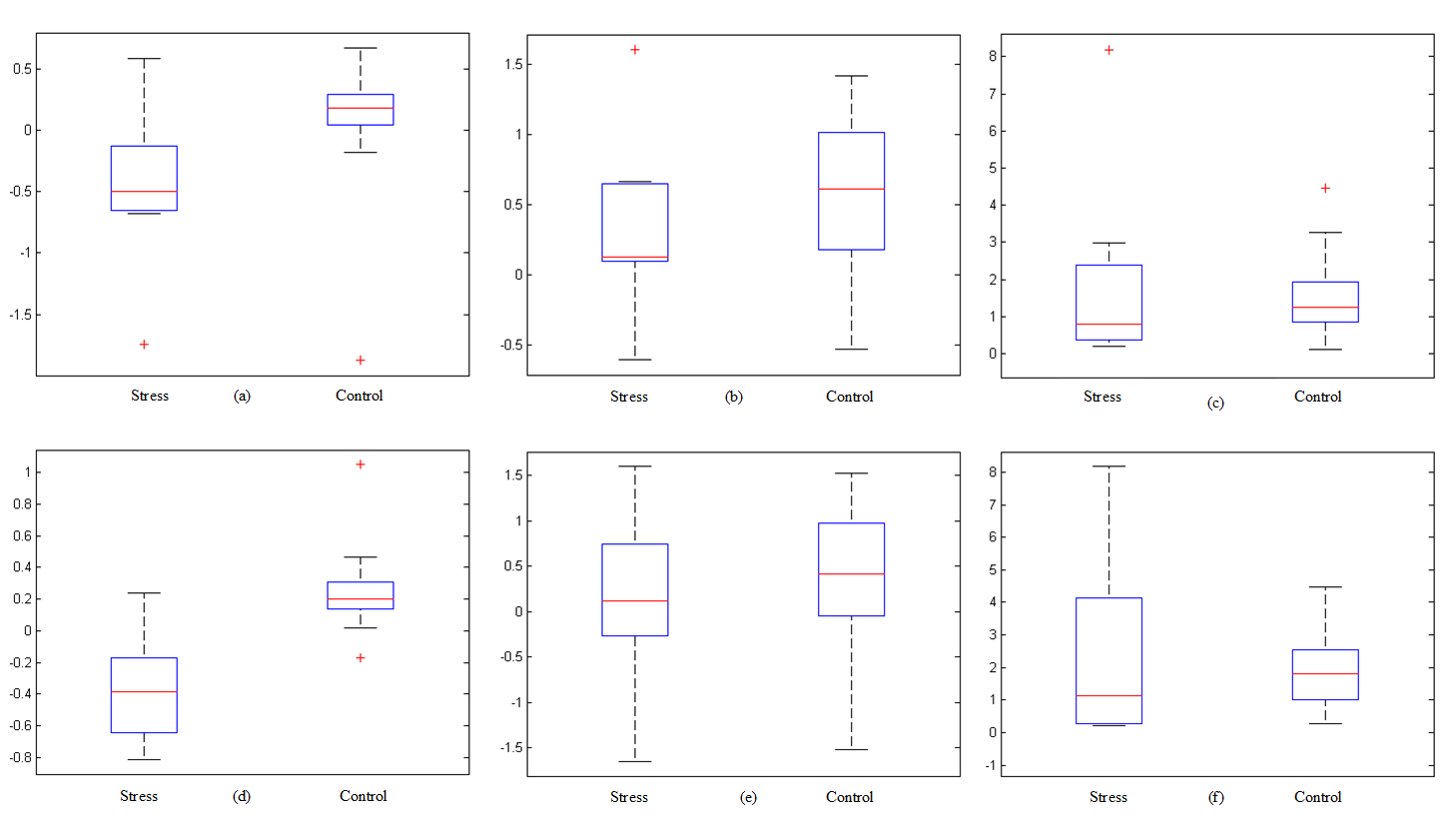}
\caption{Box plots of features (a) Alpha asymmetry (b) beta (c) gamma (d)Alpha asymmetry (EE) (e) beta (EE) (f) gamma(EE), where EE represents the labeling method of expert evaluation.}
\label{fig:boxplot}
\end{center}
\end{figure*}
A t-test was applied on the extracted features to find whether a statistically significant difference exist between the stress and control groups. Both labeling methods are considered for the t-test and results are shown in the Table \ref{PSSt} for different EEG oscillations. It is evident that at a confidence level of $0.05$, none of the extracted feature are found statistically significant in the stress and control condition, when PSS based labelling is used. It is also evident that beta and gamma waves from $AF3$ are statistically significant features in the stress and control group, when labels assigned by expert evaluation are used. Five additional features, namely frontal ($\alpha_{f}$) and temporal ($\alpha_{t}$) alpha asymmetries, frontal ($\beta_{f}$) and temporal ($\beta_{t}$) beta asymmetries, and alpha asymmetry ($\alpha_{a}$) are also used. Results of the t-test applied over these features in stress and control groups are presented in the Table \ref{tab:ratiot}. It can be seen that alpha asymmetry is statistically different in stress and control group using expert based labelling. Three significant features namely, beta (AF3), gamma (AF3) and alpha asymmetry are selected to be used in long-term stress classification based on the results of t-test.

The box plots are presented in \Fig{boxplot}, where the first row represents features acquired through PSS labeling including alpha asymmetry, beta, and gamma respectively. The second row shows the same features acquired through expert evaluation. The $+$ indicates an outlier, and the red line within the box represents the median value. %Box plots enable us to study the distribution based properties of a group. Four equal sized groups are made from the ordered data. That is, 25\% of all scores are placed in each group, referred to as quartile groups. These quartiles are numbered from 1 to 4 starting at the bottom. 
A comparatively short box plot suggests that the features are in agreement with each other. A taller box plot suggests features show different distribution within themselves. From box plots (\Fig{boxplot} (b), (c), (e), and (f)) it is observed that there is not much difference in the beta and relative gamma features to differentiate stress and control groups for both expert and PSS based labels. However, \Fig{boxplot} (a) and (d) are candidates for good features as they appear to differentiate the stress and control group. In \Fig{boxplot} (a), the alpha asymmetry for the stressed group does not have a long lower whisker, which shows alpha asymmetry is not varied along the negative quartile, while in \Fig{boxplot} (d), the stressed group has varied alpha asymmetry as shown by the lower and upper whiskers. Also, in \Fig{boxplot} (d) the median is comparatively at the center of the distribution. This suggests \Fig{boxplot} (d), which is alpha asymmetry with expert based labeling is a good candidate to be used in the stress classification task.

\begin{table}[!t]
\centering

\caption{Accuracy of classifiers for various combinations of statistically significant features.}
\label{accuracyt}
\begin{tabular}{c| ccccc}
\hline\hline
            
Features              & SVM         & NB & KNN & LR & MLP \\ \\
\hline
$\alpha_a$             & \textbf{85.20}          & 80.11 & 65.32  & 85.15 & 80.12  \\
$\gamma$              & \textbf{70.32}          & 50.21 & 50.43  & 50.33 & 50.17  \\
$\beta$               & \textbf{55.07}          & 50.01 & 50.51  & 50.48 & 50.70  \\
$\beta$, $\gamma$         & \textbf{70.45}          & 50.65 & 50.09  & 50.65 & 50.02  \\
$\alpha_a$,$\beta$        & \textbf{85.15}         & 80.02 & 65.38  & 85.04 & 85.01  \\
$\alpha_a$, $\gamma$       & 80.91          & 80.79 & 65.55  & \textbf{85.08} & 85.05  \\
$\alpha_a$, $\beta$, $\gamma$ & 80.83          & 80.77 & 65.96  & 85.09 & \textbf{85.13}  \\ 
\hline
\end{tabular}
\end{table}

\begin{table}[!t]
    %\caption{Global caption}
   %\begin{minipage}{.3\linewidth}
   %\tbl
      \caption{Evaluation parameters for the best performing classifiers with $\alpha_{a}$ as a feature.}
      \label{tab:ttc}
      \centering
      \scalebox{0.95}{
        \begin{tabular}{c |c |c |c |c |c}    % centered columns (4 columns)
\hline\hline %inserts double horizontal lines

Classifier & Average & Kappa & F-Measure & MAE  &  RMAE \\ 
& accuracy (\% ) & & & &\\
                        \hline    % inserts single horizontal line
LR	 & 85.15 & 0.70 & 0.85  & 0.22 & 0.36 \\
SVM 				 & 85.20 & 0.71 & 0.87 &  0.15 & 0.39 \\
\hline          %inserts single line
\end{tabular}}
%\end{minipage}%
\end{table}

\begin{table*}[!t]
    %\caption{Global caption}
   %\begin{minipage}{.3\linewidth}
   %\tbl
\caption{Comparison of results with previously related EEG based studies.}
\label{tab:compt}
\centering
\scalebox{1}{
\begin{tabular}{l |c c c c c}    % centered columns (4 columns)
\hline\hline %inserts double horizontal lines
 
 & \textbf{Stress Inducer} & \textbf{Participants} & \textbf{Classifier}  & \textbf{Accuracy} (\% )\\ \\ \hline
                            % inserts single horizontal line
Lin et. al. \cite{lin2008noninvasive}& Driving simulator  & 6 & KNN and \textbf{NBC} &  71.77 \\ % inserting body of the table
Saidatul et. al. \cite{vijean2011mental}& Mental arithmetic task  & 5 &NN&  91.17 \\ % inserting body of the table
Khosrowabadi et. al \cite{khosrowabadi2011brain} &Examination & 26 & KNN and SVM &  90.00 \\ % inserting body of the table
Jun et. al.\cite{jun2016eeg}& Arithmetic task and stroop  & 10 & SVM &  96.00\\ % inserting body of the table
Al Shargie et. al. \cite{al2018towards} & Mental arithmetic task & 18 & SVM and \textbf{ECoC} &  95.37 \\ % inserting body of the t
Subhani et. al. \cite{subhani2017machine}&MIST  & 42 & LR, SVM and NB &  94.60 \\ % inserting body of the table 
\hline
Sanay et. al. \cite{saeed2017quantification} &None & 28  & NB &  71.43\\
\textbf{Proposed} &\textbf{None} &\textbf{33}  &  \textbf{SVM} &  \textbf{85.20} \\ % inserting body of the table
\hline          %inserts single line
\end{tabular}}
%\end{minipage}%
\end{table*}
\subsection{Classification}
Five classifiers namely, KNN, NB, SVM, LR and MLP were used with alpha asymmetry, beta and gamma waves from channel $AF3$ as features to classify long-term stress. Each combination of the selected features is analyzed with each of the classifier. The results of these classifiers in terms of average accuracy are shown in Table \ref{accuracyt}. We observed that the classifier accuracy was high whenever alpha asymmetry was either used as a single feature or in a combination with other features. The SVM and LR based classifiers gives the highest accuracy when alpha asymmetry is used as a feature. The performance evaluation parameters for these classifiers are given in \Tab{ttc}. We also observed that both SVM and LR show very similar values for kappa statistic and F-measure. SVM may have a slightly lesser mean absolute error of $0.15$ than that of logistic regression with a value of  $0.22$, whereas, LR has a lesser RMAE of $0.36$ than that of SVM i.e., $0.38$. The overall classification accuracy of both these classifier is similar. Overall, we concluded that SVM may be a better choice for an assisting system for stress recognition.

\subsection{Discussion}\label{sec:dis}
Numerous studies have analyzed brain activities under stressful conditions, which are induced by a task such as impromptu speech, examination, mental task, public speaking and cold pressor \cite{knaus2007processing,matsunami2001generator,lewis2007effect,seo2008relation,miller1993beta,hassellund2010long}. These studies evaluate short-term induced stress whereas the classification of long-term stress using EEG has not been widely investigated. In \Tab{compt}, studies involving EEG to classify human stress are presented for comparison. It is observed that different stress inducing tasks are used such as driving simulation, examination and mental arithmetic tasks. Specialized instruments like MIST and Stroop tests are also used to induce stress. For chronic stress there could be several stressors which affect the physical, emotional, cognitive or behavioral well being of a human being. Therefore, it is proposed that recording baseline EEG for stress classification is a better choice without involving stress induction. The number of participants involved in such studies vary from $5$ to $42$. The SVM and NB are used as classifiers in most of the studies. SVM is found to be the most efficient classifier, giving a maximum accuracy of $96\%$, when stress is induced by mental arithmetic test. In \cite{peng2013method}, the baseline EEG is recorded for two minutes and a non-linear analysis is performed but no classification algorithm is used. In \cite{khosrowabadi2011brain}, chronic stress has been classified with an accuracy of $90\%$, using EEG recordings from eight electrodes and a stress inducing condition. 
It is shown in this study that the alpha asymmetry of brain can be considered as a potential bio-marker for recognition of human chronic stress. The labeling should be performed by using a hybrid method (psychology expert and PSS scores) for training the system in a supervised manner. 

\section{Conclusion}\label{sec:conc}
In this paper, two different labeling methods have been used for the classification of long-term stress in humans using EEG signals. Forty-five signal features have been analyzed for classification of chronic stress and alpha asymmetry was found to be a discriminating feature when using expert's evaluation as ground truth. The PSS scores, when used solely for labeling returned no significant features. Furthermore, it is evident from results that SVM and LR gives the highest accuracy of $85.20\%$. We also observed that the stress group was better classified as compared to control group irrespective of the classifiers used. Finally, it is established that alpha asymmetry can be used a potential bio-marker for the classification of long-term stress with SVM. To the best of our knowledge, none of the previous EEG based studies have involved a psychology expert for labeling of groups for long-term stress assessment. In future, more features and participants will be considered for the analysis. 

\bibliographystyle{IEEEtran}
\bibliography{ref.bib}

\end{document}